\documentclass[twocolumn,showpacs,amsmath,amssymb]{revtex4}

\usepackage{graphicx}

\renewcommand{\vec}[1]{\mbox{\boldmath $#1$}}

\begin{document}

\title{
Relation between shrinkage effect and compression of rotational spectrum in $^7_\Lambda$Li hypernucleus}

\author{K. Hagino and T. Koike}
\affiliation{ 
Department of Physics, Tohoku University, Sendai, 980-8578,  Japan} 


\begin{abstract}
It has been shown experimentally that the $^6$Li nucleus shrinks by adding 
a $\Lambda$ particle while its spectrum is also compressed. 
We discuss the compatibility of these two effects, contrasting also with 
a relation between the shrinkage effect and the stability of the spectrum in 
the $^9_\Lambda$Be hypernucleus. 
To this end, we employ two-body $d$-$^5_\Lambda$He and 
$\alpha$-$^5_\Lambda$He 
cluster 
models for the $^7_\Lambda$Li and $^9_\Lambda$Be hypernuclei, respectively. 
We first argue that a Gaussian-like interaction between two clusters 
leads to a stabilization of the spectrum against an addition of a $\Lambda$ particle,  
even though the intercluster distance is reduced. 
In the case of $^7_\Lambda$Li, 
the spin-orbit interaction 
between the intercluster motion and the deuteron spin has to be considered also. 
We show that the shrinkage effect makes 
the expectation value of the spin-orbit potential 
larger, lowering the excitation energy for 
the $^6$Li($3^+)\otimes \Lambda(1/2^+)$ level. 
\end{abstract}


\maketitle

\section{Introduction}

One of the main interests in physics of hypernuclei is to 
investigate the impurity effect of $\Lambda$ particle on 
the structure of atomic nuclei. 
This includes the change of {\it e.g.,} nuclear size\cite{MBI83}, the 
density distribution\cite{HKYMR10}, 
deformation properties\cite{Z80,ZSSWZ07,WH08,SWHS10,WHK11,YLHWZM11,LZZ11,IKDO11}, 
the neutron drip-line\cite{VPLR98,ZPSV08}, 
and fission barrier\cite{MCH09}. 
A characteristic feature of $\Lambda$ particle is that it is free from the Pauli principle from nucleons, 
and thus it can deeply penetrate into the nuclear interior. 
It has been predicted that the $\Lambda$ particle in the center of a nucleus attracts surrounding 
nucleons, leading to a shrinkage of nuclear size \cite{MBI83}. 

High precision $\gamma$-ray spectroscopy measurements have by now 
been systematically carried out for the p-shell $\Lambda$ hypernuclei \cite{HT06}. 
Such experiments have revealed that the electronic quadrupole transition probability, $B(E2)$, 
from the first excited state (3$^+$) to the ground state (1$^+$) of $^6$Li is considerably 
reduced when a $\Lambda$ particle is added \cite{T01}. 
By adopting a simple $\alpha$-$d$ and $^5_\Lambda$He-$d$ cluster models 
for $^6$Li and $^7_\Lambda$Li, respectively, 
the reduction in $B(E2)$ has been interpreted as a reduction of the intercluster distance 
by 19\%\cite{T01}, in accordance with theoretical calculations \cite{MBI83,HKMM99}. 

Although this interpretation has been well accepted, the spectrum of $^7_\Lambda$Li shows a somewhat 
puzzling feature. 
That is, one would naively think that the reduction of the intercluster size leads to a smaller moment of 
inertia, and thus the rotational excitation energy would increase. 
This has indeed been observed in the $^{13}_{~\Lambda}$C hypernucleus\cite{K02} (notice that only the energy of 
the 3/2$^+$ state has been measured in Ref. \cite{K02}. However, the theoretical calculation suggests that 
the partner state of the doublet (5/2$^+$) lies below the 3/2$^+$ state only by at most 
0.36 MeV \cite{HKMYY00}. This 
corresponds 
to the spin averaged energy of the doublet of 4.66 MeV, 
that is compared to the energy of the 2$^+$ state in 
$^{12}$C, 4.44 MeV). 
On the contrary, the observed difference in the spin averaged energy between the first ($1/2^+,3/2^+$) 
and the second ($5/2^+,7/2^+$) doublet levels in $^7_\Lambda$Li is 1.858 MeV \cite{U06}, that is smaller than the 
excitation energy (2.186 MeV) of the first excited state of $^6$Li (see Fig. 1). 
For the $^9_\Lambda$Be hypernucleus, the experimental data show that 
the spin averaged energy of the $^8$Be(2$^+$)$\otimes \Lambda(1/2^+)$ doublet 
states (3/2$^+$, 5/2$^+$) 
is almost the same as the energy of the first 2$^+$ state of the $^8$Be nucleus \cite{A02,T05}, despite 
a similar shrinkage effect, expected as in the $^7_\Lambda$Li nucleus\cite{MBI83}. 
The stability of the rotational energy of $^9_\Lambda$Be was shown to remain the same even when the $\Lambda$ 
particle is replaced by a charmed baryon $\Lambda_c$ \cite{BB82}. 
Apparently, these behaviors of the spectrum 
of the $^7_\Lambda$Li and $^9_\Lambda$Be cannot be understood in analogy of 
{\it e.g.,} a classical rotor of diatomic molecules. 
We summarize the experimental spaectra in Fig. 1. 

\begin{figure}[htb]
\includegraphics[scale=0.35,clip]{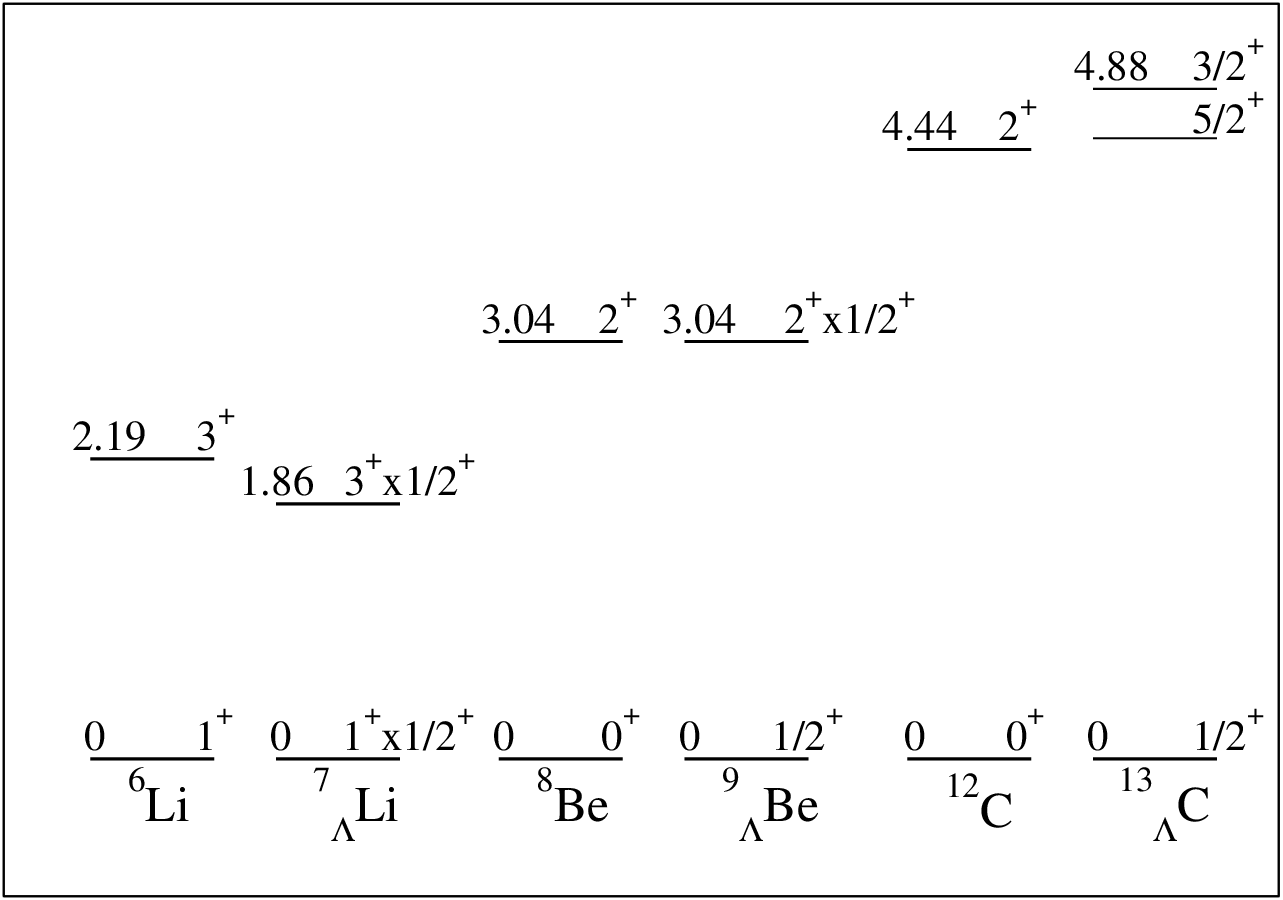}
\caption{The experimental low-lying spectra for 
$^6$Li, $^7_\Lambda$Li, $^8$Be, $^9_\Lambda$Be, $^{12}$C, and $^{13}_{~\Lambda}$C nuclei. 
The energies are denoted in the units of MeV. 
The energies for the 1$^+\otimes$1/2$^+$ and 2$^+\otimes$1/2$^+$ levels of the 
$^7_\Lambda$Li nucleus are obtained by spin-averaging the observed level energies for 
the 1/2$^+$ (0 MeV) and the 3/2$^+$ (0.069 MeV) states,  
and the 5/2$^+$ (2.05 MeV) and the 7/2$^+$ (2.52 MeV) states\cite{U06}, respectively. 
The energy for the 2$^+\otimes$1/2$^+$ level of the $^9_\Lambda$Be is the spin averaged 
energy between the 5/2$^+$ (3.024 MeV) and 3/2$^+$ (3.067 MeV) states \cite{T05}. 
In the $^{13}_{~\Lambda}$C nucleus, only the energy for the 3/2$^+$ state has been measured \cite{K02}, 
while the energy for the 5/2$^+$ state is expected to be lower than the energy of the 3/2$^+$ state 
by at most 0.36 MeV \cite{HKMYY00}. }
\end{figure}

In this paper, we clarify the relation between the two contradictory effects of $\Lambda$ particle on 
the structure of light atomic nuclei, that is, the shrinkage of the size and the compression/stabilization of 
the spectrum.  
An important point to remember 
is that the $\alpha$ cluster structure is well developed in the ground state of $^6$Li and 
$^8$Be, while the ground state takes the mean-field-like structure in $^{12}$C and 
in heavier nuclei. Our aim in this paper is thus to discuss how the spectra of 
$^7_\Lambda$Li and $^9_\Lambda$Be are compatible with the shrinkage of their size from a viewpoint of 
a two-body cluster model. 

The paper is organized as follows. In Sec. II, we first discuss the impurity effect of $\Lambda$ in the 
$^9_\Lambda$Be nucleus. Using the 
$\alpha$-$^5_\Lambda$He cluster model together with the semi-classical Bohr-Sommerfeld quantization rule, 
we investigate how a $\Lambda$ particle influences the rotational spectrum. 
In Sec. III, we discuss the rotational spectrum of the 
$^7_\Lambda$Li nucleus. We show that the spin-orbit interaction due to the deuteron spin plays an 
important role in lowering the excitation energy of the $^6$Li$\otimes\Lambda(1/2^+)$ level. 
We then summarize the paper in Sec. IV. 

\section{stabilization of level scheme for $^9_\Lambda$Be}

We first discuss the level scheme of the $^9_\Lambda$Be hypernucleus, and clarify the 
reasoning why the influence of the $\Lambda$ particle in the level spacing is negligibly small. 
For this purpose, we use a simple two-body cluster model with $\alpha$-$^5_\Lambda$He configuration. 
Assuming a local potential, an intercluster potential at an intercluster distance $R$ 
reads
\begin{equation}
V_{\alpha^5_\Lambda {\rm He}}(R)=V^{(N)}_{\alpha\alpha}(R)+V^{(C)}_{\alpha\alpha}(R)
+V_{\alpha\Lambda}(R), 
\end{equation}
where 
$V^{(N)}_{\alpha\alpha}$ and $V^{(C)}_{\alpha\alpha}$ are the nuclear and the Coulomb parts of the 
potential between the $\alpha$ particle and the core nucleus ($^4$He) of the $^5_\Lambda$He, and 
$V_{\alpha\Lambda}$ is the potential between the $\Lambda$ particle in $^5_\Lambda$He and 
the $\alpha$ particle. 
Assuming that the center of mass of $^5_\Lambda$He is identical to the center 
of mass of the core nucleus $^4$He, we use the same potential as that given in Ref. \cite{BFW77} for 
$V^{(N)}_{\alpha\alpha}$ and $V^{(C)}_{\alpha\alpha}$. 
That is, 
\begin{equation}
V^{(N)}_{\alpha\alpha}(R)=-V_0\,e^{-\alpha R^2}, 
\label{gauss_aa}
\end{equation}
with $V_0$=122.6225 MeV and $\alpha$=0.22 fm$^{-2}$, and 
\begin{equation}
V^{(C)}_{\alpha\alpha}(R)=\frac{4e^2}{R}\,{\rm erf}(\beta R^2),
\end{equation}
with $\beta$=0.75 fm$^{-1}$, where erf$(x)$ is the error function. 
For $V_{\alpha\Lambda}$, we assume that it is given by the direct part of the 
double folding potential, that is, 
\begin{equation}
V_{\alpha\Lambda}(R)=
\int d\vec{r}_Nd\vec{r}_\Lambda\,
\rho_\alpha(\vec{r}_N)
\rho_\Lambda(\vec{r}_\Lambda)
v_{N\Lambda}(\vec{R}+\vec{r}_N-\vec{r}_\Lambda), 
\label{dfm}
\end{equation}
where 
$v_{N\Lambda}$ is a nucleon-$\Lambda$ particle interaction, while 
$\rho_\alpha$ and $\rho_\Lambda$ are the density distributions for the $\alpha$ particle and 
the $\Lambda$ particle in $^5_\Lambda$He, respectively. 
We assume Gaussian density distributions for $\rho_\alpha$ and $\rho_\Lambda$, 
\begin{eqnarray}
\rho_\alpha(r)&=&4(\pi b_\alpha)^{-3/2}e^{-r^2/b_\alpha^2}, \\
\rho_\Lambda(r)&=&(\pi b_\Lambda)^{-3/2}e^{-r^2/b_\Lambda^2}. 
\end{eqnarray}
Following Ref. \cite{MBI83}, we take $b_\alpha=1.358$ fm, and 
$b_\Lambda=\sqrt{(4M_N+M_\Lambda)/4M_\Lambda}\,b_\alpha$, where $M_N$ and $M_\Lambda$ are the 
mass of nucleon and $\Lambda$ particle, respectively. 
For $v_{N\Lambda}$, we take the central part of the potential given in Ref. \cite{MBI83}, 
\begin{equation}
v_{N\Lambda}(r)=v_0\,e^{-r^2/b_v^2},
\end{equation}
with $b_v=1.034$ fm, but we adjust the strength $v_0$ so as to reproduce the 
energy of the ground state of $^9_\Lambda$Be from the threshold of the $\alpha+^5_\Lambda$He 
configuration (that is, $-$3.50 MeV). 
Since $\rho_\alpha$, $\rho_\Lambda$, and $v_{N\Lambda}$ are all given in a Gaussian form, 
the double folding integral in Eq. (\ref{dfm}) can be evaluated analytically as 
\begin{equation}
V_{\alpha\Lambda}(R)=
4v_0\left(\frac{b_v^2}{b_\Lambda^2+b_\alpha^2+b_v^2}\right)^{3/2}
\,\exp\left(-\frac{r^2}{b_\Lambda^2+b_\alpha^2+b_v^2}\right). 
\end{equation}

With the intercluster potential so constructed, we solve 
the Schr\"odinger equation for the relative motion for each angular momentum $L$, 
\begin{equation}
\left(-\frac{\hbar^2}{2\mu_{\alpha^5_\Lambda{\rm He}}}\vec{\nabla}^2
+V_{\alpha^5_\Lambda{\rm He}}(R)-E\right)\psi_L(\vec{R})=0, 
\end{equation}
where 
$\mu_{\alpha^5_\Lambda{\rm He}}=4M_N(4M_N+m_\Lambda)/(8M_N+M_\Lambda)$ is the reduced mass for the 
relative motion between $\alpha$ and $^5_\Lambda$He. 
The Pauli principle is taken into account by excluding those states which satisfy 
$2n+L<4$ \cite{BFW77}, where $n$ is the radial node of the wave function. 

The E2 transition probability can be computed with the intercluster wave functions $\psi_L(\vec{R})$. 
In the two-body cluster model of $(A_1,Z_1)+(A_2,Z_2)$, the E2 transition operator reads
\begin{equation}
\hat{T}_{\rm E2} = e_{\rm E2}R^2Y_{2\mu}(\hat{\vec{R}}),
\end{equation}
where 
\begin{equation}
e_{\rm E2}=\left(\frac{M_1}{M_1+M_2}\right)^2Z_2+\left(\frac{M_2}{M_1+M_2}\right)^2Z_1,
\end{equation}
is the E2 effective charge, $M_1$ and $M_2$ being the mass of the fragments 1 and 2, respectively. 

\begin{figure}[htb]
\includegraphics[scale=0.5,clip]{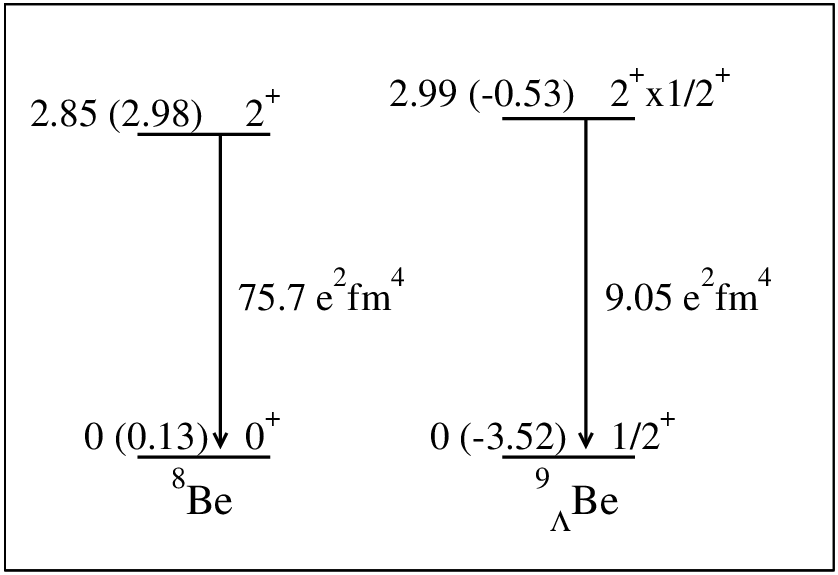}
\caption{
Low-lying spectra for 
$^8$Be and $^9_\Lambda$Be
obtained with the two-body cluster model of $\alpha$+$\alpha$ and 
$\alpha+^5_\Lambda$He, respectively. 
The energies are denoted in the units of MeV. 
The energies measured from the two-body thresholds are also shown in the parentheses. 
The arrows indicate the reduced $E2$ transition probabilities, $B(E2)$, 
from the first excited state to the ground state.  }
\end{figure}

The calculated spectra for the $^8$Be and $^9_\Lambda$Be are shown in Fig. 2. 
Since we do not include the spin-dependent part of nucleon-$\Lambda$ interaction, the 
doublet states (3/2$^+$,5/2$^+$) of $^8$Be(2$^+)\otimes\Lambda(1/2^+)$ are degenerate in energy. 
In order to obtain the energy and the wave function for resonance states, we use 
a bound state approximation. That is, we have replaced the intercluster potential 
for $R$ larger than the barrier radius with a constant so that the wave functions are 
confined inside the potential barrier\cite{JRB77,GK87,G88}. 
We have confirmed that the energies obtained with this procedure are close to those energies 
that give the maximum of the energy derivative of the scattering phase shift. 
For instance, for the 2$^+$ state of $^8$Be, the energy derivative of the phase shift is 
maximum at $E$=2.80 MeV, while the energy obtained in the bound state approximation is 2.85 MeV. 

\begin{figure}[htb]
\includegraphics[scale=0.5,clip]{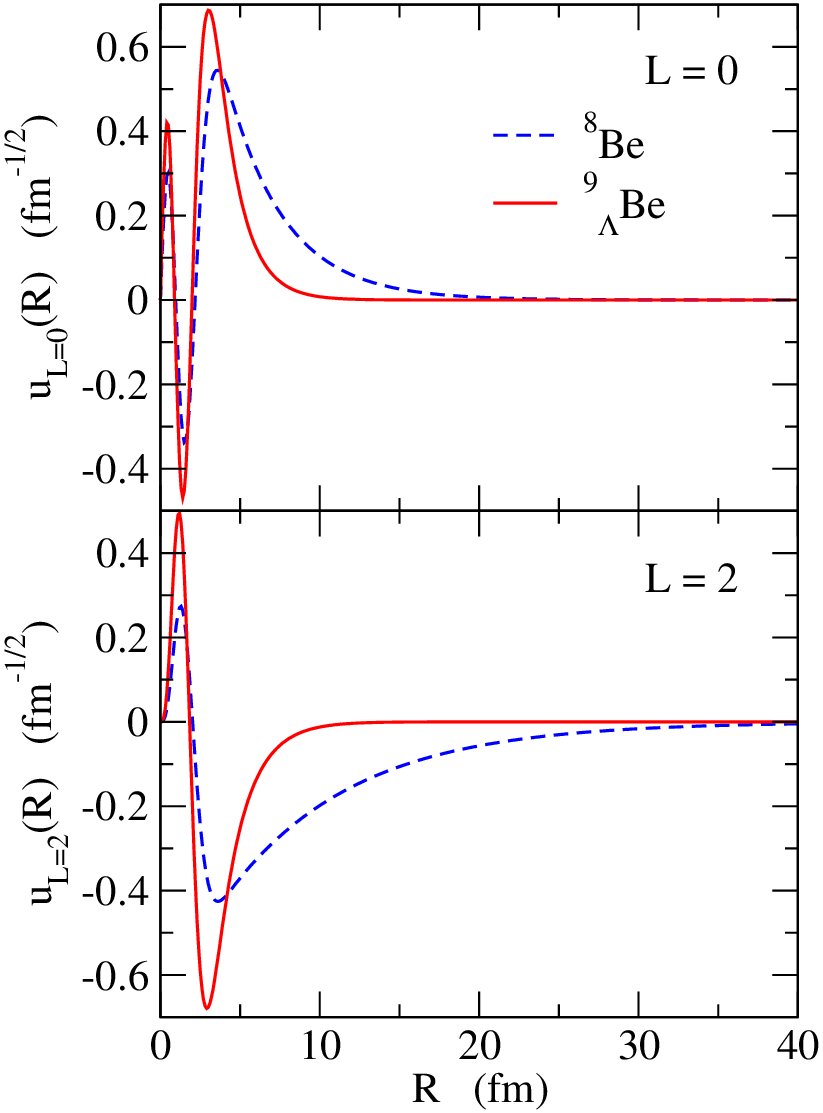}
\caption{(Color online)
The radial wave functions $u_L(R)$ defined as $\psi_L(\vec{R})=u_L(R)/R\cdot Y_{LM}(\hat{\vec{R}})$ 
obtained with the two-body cluster model. 
The dashed and the solid lines are for 
the $^8$Be and $^9_\Lambda$Be nuclei, respectively. 
The upper panel shows the wave functions for the ground state with $L=0$, while the lower panel 
shows the wave functions for the first excited state with $L=2$. }
\end{figure}

From Fig. 2, one sees that the excitation energy slightly increases 
due to the addition of $\Lambda$ particle. 
We have confirmed numerically that the potential which we use well reproduces the experimental 
phase shift for $\alpha$+$\alpha$ scattering\cite{BFW77}. 
If the resonance states were properly analysed, this potential would therefore yield the excitation 
energy of the 2$^+$ state in $^8$Be of 3.04 MeV (see Fig. 1). Then, the energy shift due to 
the addition of $\Lambda$ particle is estimated to be small indeed, 
that is, as small as 0.05 MeV. 
In contrast to the excitation energies, 
the B(E2) value from the first excited state to the ground state is altered drastically. 
The large change of $B(E2)$ value is due to the fact that the attraction of the $\Lambda$ particle makes the resonance 
levels of $^8$Be turn to the bound states in $^9_\Lambda$Be. 
As a consequence, the wave functions become spatially much more compact, as shown in Fig. 3. 
The root-mean-square distance between the fragments is 4.90 fm and 3.24 fm for the $L=0$ states of 
$^8$Be and $^9_\Lambda$Be, respectively. The energy change due to the $\Lambda$ particle 
shown in Fig. 2, that is, 0.14 MeV, is much smaller 
than what would have been expected from a classical rotor, $E_L=L(L+1)\hbar^2/2\mu R^2$, which corresponds to 
the energy shift of 2.64 MeV. 
Notice that the B(E2) value for the transition from the 2$^+$ to the 0$^+$ states in $^8$Be 
is consistent 
with the value obtained in Ref. \cite{LR86} using a damping factor method to evaluate the integral with 
the resonance wave functions, although this value is rather large as compared with that obtained 
in Ref. \cite{MBI83} using the harmonic oscillator expansion of the wave functions. 

\begin{figure}[htb]
\includegraphics[scale=0.5,clip]{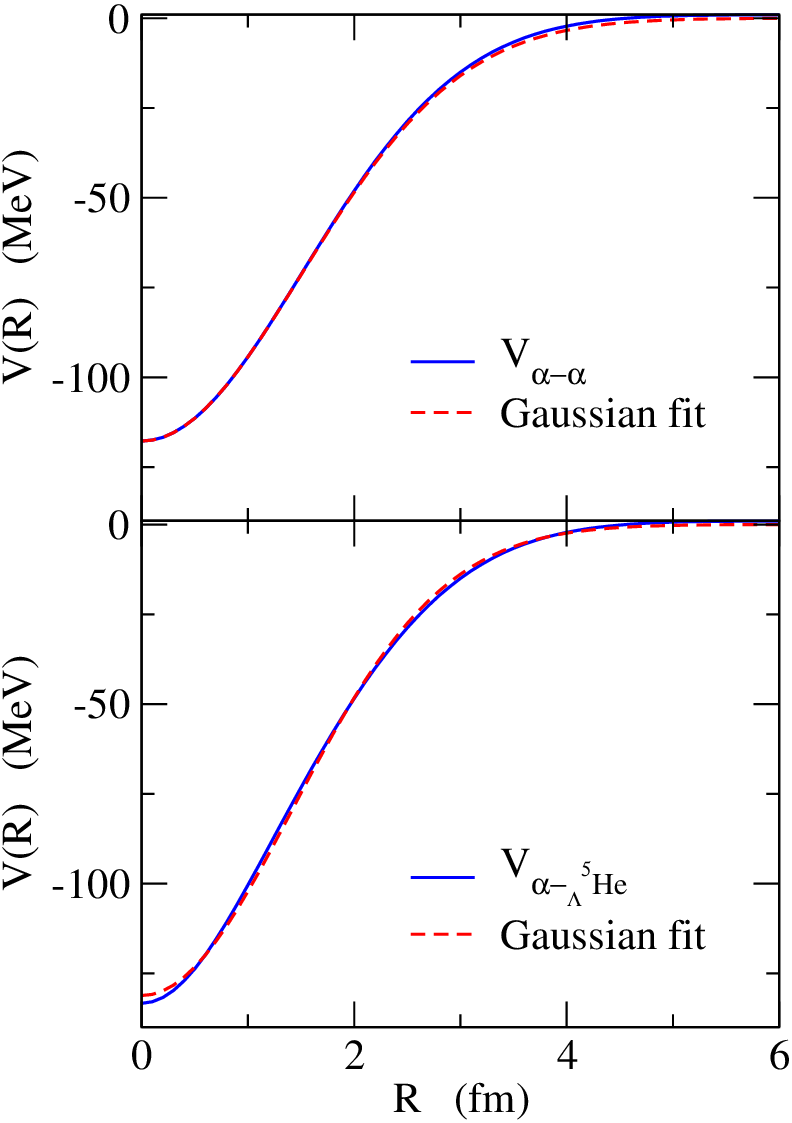}
\caption{(Color online)
The intercluster potential for 
the $\alpha+\alpha$ system (the upper panel) and 
for the $\alpha+^5_\Lambda$He system (the lower panel). 
The dashed lines show the result of the fitting with a single Gaussian function. }
\end{figure}

We now ask a question why the spectrum of $^9_\Lambda$Be is not influenced much, despite the fact that  
the wave functions are considerably altered. In order to address this question, the semi-classical 
approximation may be useful. In Ref. \cite{R77}, Rowley used the Bohr-Sommerfeld quantization 
rule to show that the spectrum is approximately given by 
\begin{equation}
E_L-E_{L=0}\sim \frac{\beta\hbar^2}{8\mu}\,L(L+1),
\label{rowley}
\end{equation}
for a Gaussian intercluster potential, $V(R)=V_0\,e^{-\beta R^2}$. 
We numerically confirm this relation in Appendix A. 
This formula indicates that the spectrum depends 
on neither the depth parameter $V_0$ nor the intercluster distance, 
but 
only on the range parameter $\beta$ of the potential 
as well as the reduced mass $\mu$. 

We show the intercluster potentials for the present problem in Fig. 4 by the solid lines.  
The potential for the $\alpha+^5_\Lambda$He is deeper than that for $\alpha+\alpha$ by 
15.6 MeV, due to an additional attraction caused by the $\Lambda$ particle. 
These potentials are actually well fitted with the Gaussian form 
with $V_0=-117.8$ MeV and 
$\beta=0.222$ fm$^{-2}$ for $^8$Be and 
with $V_0=-131.2$ MeV and 
$\beta=0.250$ fm$^{-2}$ for $^9_\Lambda$Be. 
The quality of the fit can be seen in Fig. 4. 
The reduced masses, on the other hand, are 
1877.8 MeV and 2120.6 MeV for the 
$\alpha+\alpha$ and 
the $\alpha+^5_\Lambda$He systems, respectively. Here, we have used the average value 
of the proton and neutron masses as the nucleon mass. 
The factor 
$\beta\hbar^2/8\mu$ in Eq. (\ref{rowley}) is thus 0.574 MeV for the 
$\alpha+\alpha$ system and 0.572 MeV for the 
the $\alpha+^5_\Lambda$He system. 
It is remarkable that these values are considerably close to each other. 
Evidently, a Gaussian-like intercluster potential has a responsibility to make 
the spectra resemble to each other between $^8$Be and $^9_\Lambda$Be, despite 
that the absolute value of the energies and thus the radial dependence of the 
wave functions are considerably changed due to an addition of a $\Lambda$ particle. 

\section{Level scheme for $^7_\Lambda$Li: role of spin-orbit interaction}

Let us discuss next the spectrum of the $^7_\Lambda$Li nucleus. 
If we assume that the nucleus $^6$Li takes the $\alpha$+deuteron ($d$) structure, 
a big difference of this nucleus from $^8$Be=$\alpha+\alpha$ 
is that one of the fragments ({\it i.e.,} the deuteron) has a finite spin ($S=1$). 
The potential between $\alpha$ and $d$ then reads \cite{NTJ83,NTJK84,MR85}, 
\begin{equation}
V_{\alpha d}(R)=V_0(R)+V_1(R)\vec{L}\cdot\vec{S}+
V_2(R)\left[\frac{(\vec{S}\cdot\vec{R})^2}{R^2}-\frac{1}{3}\vec{S}^2\right]. 
\end{equation}
The spectrum of the ground rotational band of $^6$Li can be understood as follows\cite{MR85}. 
This band arises from the states with $2n+L=2$, that is, $L=0$ (1s) and $L=2$ (0d). 
By combining with the deuteron spin $S=1$, the 1s state yields the total spin and parity of $I^\pi$=1$^+$, 
while the 0d state yields $I^\pi=1^+,2^+$, and $3^+$ triplet states. These triplet states are degenerate 
in energy in the 
central potential $V_0(R)$, but they split by the spin-orbit interaction 
$V_1(R)\vec{L}\cdot\vec{S}$ as well as by the tensor interaction 
$V_2(R)\left[(\vec{S}\cdot\hat{\vec{R}})^2-\vec{S}^2/3\right]$. 

In order to investigate how the spin-orbit intercluster potential affects the spectrum 
of the $^7_\Lambda$Li nucleus, we closely follow Ref. \cite{MR85} and use the semi-microscopic 
cluster model of Buck {\it et al.}\cite{BMR79}. To do so, we assume that 
the $^7_\Lambda$Li nucleus takes the $d+^5_\Lambda$He structure. 
In this model, the intercluster potentials $V_0(R), V_1(R)$, and $V_2(R)$ are constructed based on 
a core+$p$+$n$ three-body model. We regard the $\alpha$ particle and $^5_\Lambda$He as the core nucleus 
for $^6$Li and $^7_\Lambda$Li, respectively. 
In the following, for simplicity, 
we assume that the deuteron is in a pure s-state, and thus the tensor part of the intercluster potential 
vanishes. This term plays an essential role in reproducing the quadrupole moment of $^6$Li\cite{NTJ83,NTJK84,MR85}, 
but its influence 
is expected to be small for the spectrum. 
With this approximation, the central and the spin-orbit potentials read \cite{MR85}
\begin{eqnarray}
V_0(R)&=&-2\int dr\, U_0(r,R)\chi(r)^2, \\
V_1(R)&=&-\frac{2M_N}{2(M_N+M_c)} \nonumber \\
&&\times \int dr\,\left[W_0(r,R)+\frac{r}{2R}W_1(r,R)\right]\chi(r)^2, 
\label{LS}
\end{eqnarray}
where $M_c$ is the mass of the core nucleus ({\it i.e.,} $^4$He or $^5_\Lambda$He) 
and $\chi(r)$ is a s-state wave function for relative 
motion between $p$ and $n$ in the deuteron. 
$U_0, W_0$ and $W_1$ are multipole components of the nucleon-core interaction, 
\begin{equation}
V_{cN}=-U(\vec{R}\pm\vec{r}/2)-W(\vec{R}\pm\vec{r}/2)\,\vec{l}\cdot\vec{s},
\end{equation}
where $\vec{r}$ is the relative distance between $p$ and $n$ in the deuteron, $\vec{l}$ is the relative 
angular momentum between the core and the nucleon, and $\vec{s}$ is the nucleon spin. 
That is, 
$U_0, W_0$ and $W_1$ are defined as 
\begin{eqnarray}
U(\vec{R}\pm\vec{r}/2)&=&4\pi\sum_{\lambda}U_\lambda(R,r)Y_\lambda(\pm\hat{\vec{r}})\cdot Y_\lambda(\hat{\vec{R}}), \\
W(\vec{R}\pm\vec{r}/2)&=&4\pi\sum_{\lambda}W_\lambda(R,r)Y_\lambda(\pm\hat{\vec{r}})\cdot Y_\lambda(\hat{\vec{R}}).
\end{eqnarray}

In the calculations shown below, following Ref. \cite{MR85}, 
we employ the Hulthen form of the deuteron wave function, that is, 
$\chi(r)=\sqrt{2\alpha}\,e^{-\alpha r}$ with $\alpha$=0.2316 fm$^{-1}$, 
and the same potential as in Ref. \cite{MR85} (with a slight readjustment for the 
depth parameters) for the nucleon-$\alpha$ potential. 
The contribution of the $\Lambda$ particle to the $^5_\Lambda$He-nucleon potential is 
estimated with a single folding potential, 
\begin{equation}
U_{N\Lambda}(r)=
\int d\vec{r}_\Lambda\,
\rho_\Lambda(\vec{r}_\Lambda)
v_{N\Lambda}(\vec{r}+\vec{r}_N-\vec{r}_\Lambda), 
\end{equation}
where the $\Lambda$ particle density 
$\rho_\Lambda$ and the nucleon-$\Lambda$ interaction 
$v_{N\Lambda}$ are given in the previous section. 
As in Eq. (\ref{dfm}), this potential can be computed analytically as 
\begin{equation}
U_{N\Lambda}(r)=
v_0\left(\frac{b_v^2}{b_\Lambda^2+b_v^2}\right)^{3/2}
\,\exp\left(-\frac{r^2}{b_\Lambda^2+b_v^2}\right). 
\end{equation}
We adjust the value of $v_0$ in order to reproduce the ground state energy of the $^7_\Lambda$Li nucleus (that is, 
$-$3.48 MeV from the threshold of $^5_\Lambda$He+$d$). 
The total central potential for $^5_\Lambda$He-nucleon is given by 
$U(r)=U_{N\alpha}(r)+U_{N\Lambda}(r)$, where $U_{N\alpha}(r)$ is the central part of the 
nucleon-$\alpha$ potential. 
Since we neglect the spin-orbit interaction in the nucleon-$\Lambda$ interaction, the $\Lambda$ particle 
does not contribute to the spin-orbit part of the intercluster potential, $V_1(R)$, except for the trivial 
mass factor in Eq. (\ref{LS}). 

\begin{figure}[htb]
\includegraphics[scale=0.5,clip]{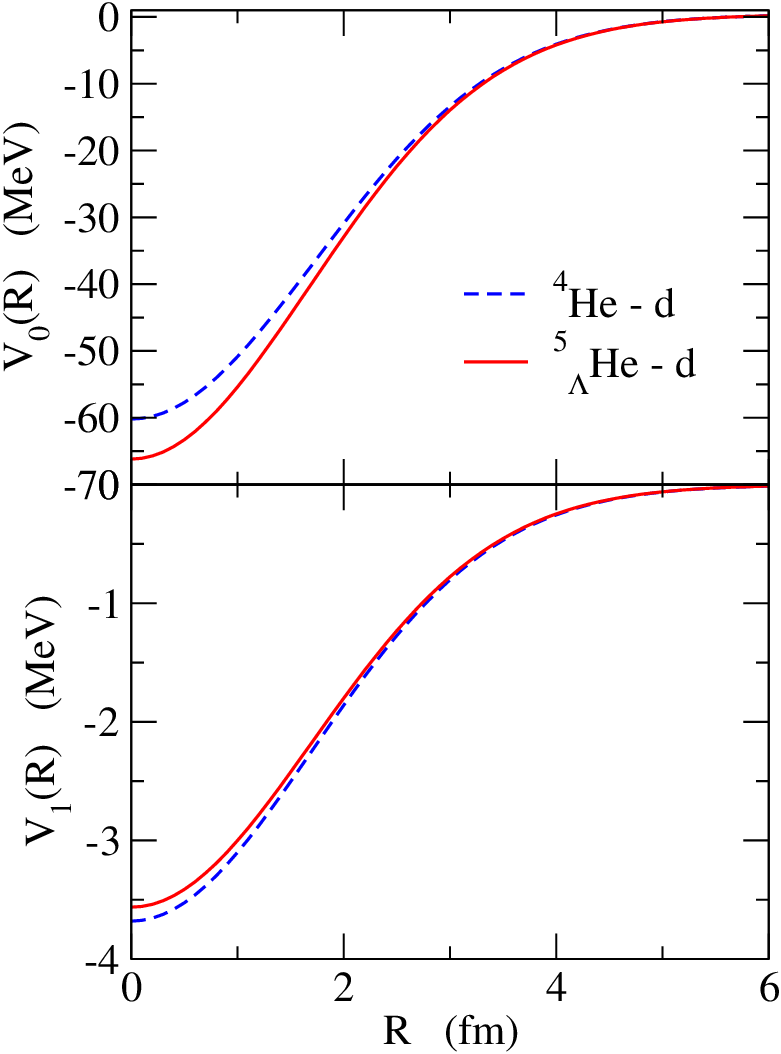}
\caption{(Color online)
The central part (the upper panel) and the spin-orbit part (the lower panel) of the intercluster potential. 
The dashed lines show the potentials for the $^6$Li nucleus, while the solid lines show those for the $^7_\Lambda$Li 
hypernucleus. }
\end{figure}

\begin{figure}[htb]
\includegraphics[scale=0.5,clip]{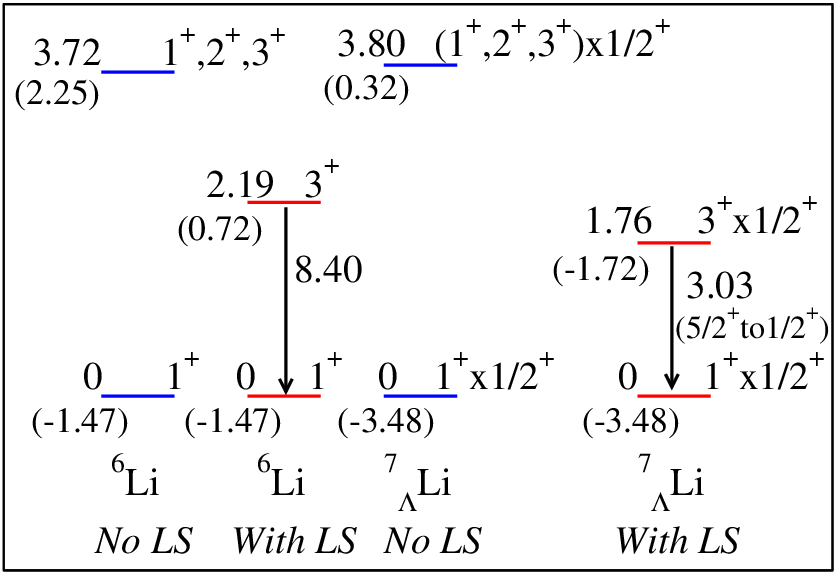}
\caption{(Color online)
Low-lying spectra for 
$^6$Li and $^7_\Lambda$Li
obtained with the two-body cluster model of $\alpha$+$d$ and 
$^5_\Lambda$He+$d$, respectively. 
The results obtained with and without the spin-obit intercluster potential are compared. 
The energies are denoted in the units of MeV. 
The energies measured from the two-body thresholds are also shown in the parentheses. 
The arrows indicate the reduced $E2$ transition probabilities, $B(E2)$, 
from the first excited state to the ground state, where the calculated values are denoted in 
the units of $e^2$fm$^4$.}
\end{figure}

\begin{figure}[htb]
\includegraphics[scale=0.45,clip]{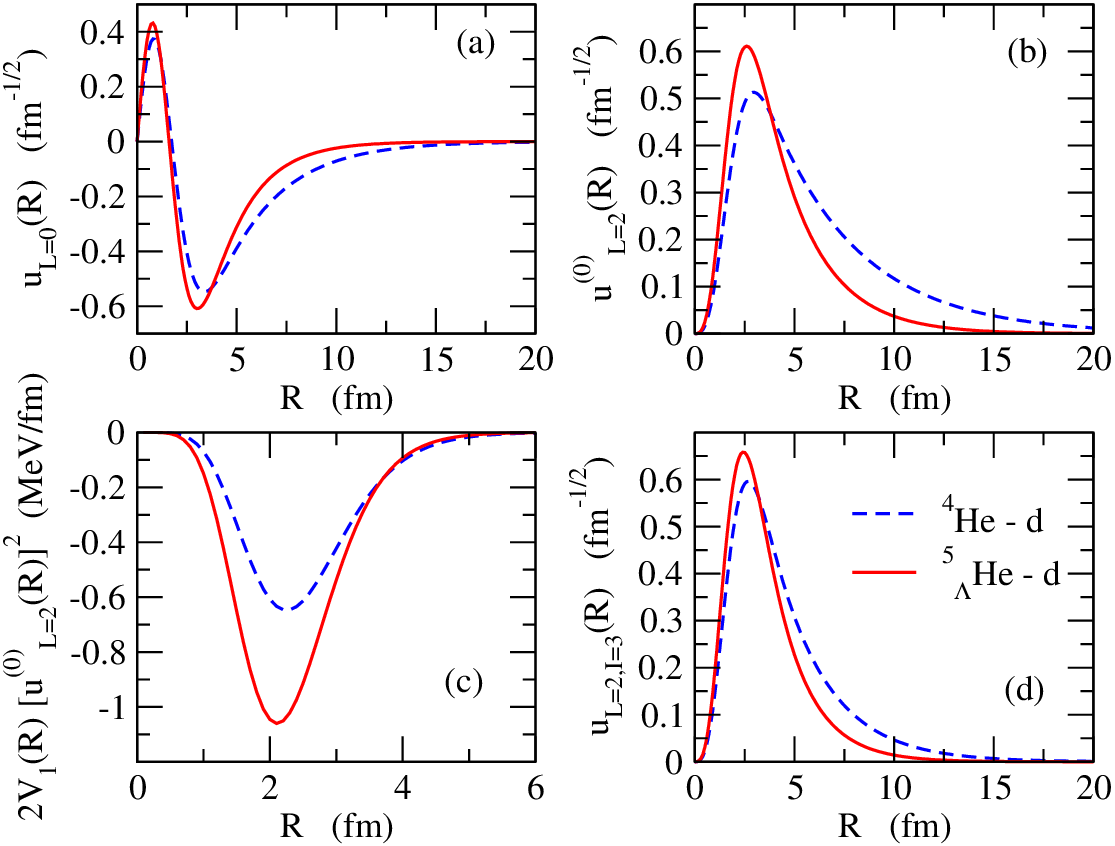}
\caption{(Color online)
(a) The radial wave functions for the ground state with $L=0$ for 
$^6$Li (the dashed line) and $^7_\Lambda$Li (the solid line). 
(b) Those for the first excited state with $L=2$ obtained without including the spin-orbit 
intercluster potential. 
(c) The overlap between the spin-orbit potential and the unperturbed wave functions for $L=2$. 
(d) The same as Fig. 7(b), but those obtained with including the spin-orbit potential. }
\end{figure}

The central part $V_0(R)$ and the spin-orbit part $V_1(R)$ of the intercluster potentials are 
shown in Fig. 5. 
For the central part, we also include the Coulomb potential evaluated in the same way as in 
Ref. \cite{MR85}. 
The solid and the dashed lines are the potentials for the $^5_\Lambda$He-$d$ and the $\alpha$-$d$, 
respectively. 
The depth of the central part increases by 5.98 MeV due to an addition of a $\Lambda$ particle, while the change of 
the spin-orbit potential is much smaller. 
The spectra obtained with these potentials are shown in Fig. 6. In the absence of the spin-orbit potential, $V_1(R)$, 
an addition of the $\Lambda$ particle slightly increases the energy of the first excited state, from 3.72 MeV 
in $^6$Li to 3.80 MeV in $^7_\Lambda$Li. This behavior is similar to the change of the spectrum of $^8$Be shown in 
Fig. 2. The radial wave functions for the ground state and the first excited state obtained without 
the spin-orbit potential are shown in Figs. 7(a) and 7(b), respectively. 
As in the case of $^9_\Lambda$Be, the shrinkage effect of the $\Lambda$ particle is substantial \cite{MBI83,HKMM99}. 

Let us now consider the effect of the spin-orbit potential. 
We fist treat it by the first order perturbation theory, that is, the energy shift for 
$L=2$ and $I=3$ is approximately 
given by 
\begin{equation}
\Delta E_{L=2,I=3}\sim 2\int dR V_1(R)\left[u_{L=2}^{(0)}(R)\right]^2.
\label{pert}
\end{equation}
Here, the factor 2 is the eigenvalue of the operator $\vec{L}\cdot\vec{S}$ for $L=2$ and $I=3$, and 
$u_{L=2}^{(0)}$ is the unperturbed wave function for $L=2$ obtained without the spin-orbit potential. 
Using the wave functions shown in Fig. 7(b), we obtain 
$\Delta E_{L=2,I=3}=-1.22$ MeV for $^6$Li and $-1.84$ MeV for $^7_\Lambda$Li. Fig. 7(c) show the integrand 
of Eq. (\ref{pert}). One can clearly see that the overlap of the wave function with the spin-orbit 
potential shown in the lower panel of Fig. 5 increases significantly for the $^7_\Lambda$Li hypernucleus due to 
the shrinkage effect, leading to the larger value of the energy shift 
$\Delta E_{L=2,I=3}$. The spectra obtained by treating the spin-orbit potential exactly, and 
the corresponding wave functions are shown in Figs. 6 and 7(d), respectively. 
The effect of the spin-orbit potential is indeed much larger in $^7_\Lambda$Li due to the shrinkage effect, 
and the energy of the $^6$Li(3$^+)\otimes\Lambda(1/2^+)$ state appears lower than that of the $^6$Li(3$^+$) state. 
We thus conclude that the spin-orbit potential due to the finite deuteron spin plays an important role in 
compressing the spectrum of the $^7_\Lambda$Li hypernucleus. 
We mention that the calculated $B(E2)$ value from the 5/2$^+$ member of the 
$^6$Li(3$^+)\otimes\Lambda(1/2^+)$ state to the 1/2$^+$ state in the 
ground state $^6$Li(1$^+)\otimes\Lambda(1/2^+)$ doublet is 3.03 $e^2$fm$^4$, that is in good agreement with 
the experimental value, 3.6$\pm$0.5$^{+0.5}_{-0.4}$ $e^2$fm$^4$ \cite{T01}. 
The calculation also reproduces well the 
experimental value for the 
energy of the $^6$Li(3$^+)\otimes\Lambda(1/2^+)$ state, 1.86 MeV (See Fig. 1). 

\section{Summary}

We have investigated the structure of $^9_\Lambda$Be and $^7_\Lambda$Li hypernuclei using two-body cluster 
models of $\alpha+d$ and $^5_\Lambda$He+$d$, respectively. We particularly discussed the relation between 
the shrinkage 
effect and the spectra of these hypernuclei.  
We first showed that the two-body cluster model yields only a small change in the spectra between $^8$Be and 
$^9_\Lambda$Be even though the intercluster distance is significantly reduced due to an addition of a 
$\Lambda$ particle. We argued based on the Bohr-Sommerfeld quantization rule that this is caused by the fact 
that intercluster potentials can be well fitted with a single Gaussian function, for which the spectra 
are approximately 
independent of the depth of the potential. A similar effect would be expected also for 
$^7_\Lambda$Li 
if the deuteron 
were a spin-less particle. In reality, a deuteron has a finite spin ($S$=1), and the spin-orbit potential is 
present in the intercluster potential. 
We have shown that the shrinkage effect leads to a large overlap between the wave function for the excited 
state and the spin-orbit potential for the $^7_\Lambda$Li hypernucleus as compared with $^6$Li. 
As a consequence, the spin-orbit potential acts effectively larger in 
$^7_\Lambda$Li, eventually leading to the compression of the spectrum. 

These behaviors of the spectra are peculiarities of a two-body cluster structure. In this sense, 
$^7_\Lambda$Li and $^9_\Lambda$Be are exceptional cases, as the ground states exhibit a well-developed 
$\alpha$ cluster structure. For heavier nuclei, the cluster structure 
appears in excited states while the ground state takes a mean-field-like configuration. 
In such situations, the shrinkage of radius would push up the spectrum, as has been seen experimentally 
in $^{13}_{~\Lambda}$C. The deformation degree of freedom also comes into a play there. 
An interesting case for a future investigation is the hypernucleus $^{19}_{~\Lambda}$F, in which 
the mean-field structure may be mixed with the $^{17}_{~\Lambda}$O+$d$ cluster structure. 
A $\gamma$-ray spectroscopy measurement has been planed on this hypernucleus at the J-PARC 
facility \cite{T11, K08}.  This will be the first for the sd-shell hypernuclei. The ground and the first 
excited states of the core $^{18}$F nucleus has a spin and parity of 1$^+$ and 3$^+$, respectively, 
which are the same for $^{6}$Li. Thus the measurement of the $^{18}$F(3$^+$)$\otimes\Lambda$(1/2$^+$) 
level will provide a strong stimulus to further studies on the sd-shell hypernuclei and beyond. 

\begin{figure}[htb]
\includegraphics[scale=0.5,clip]{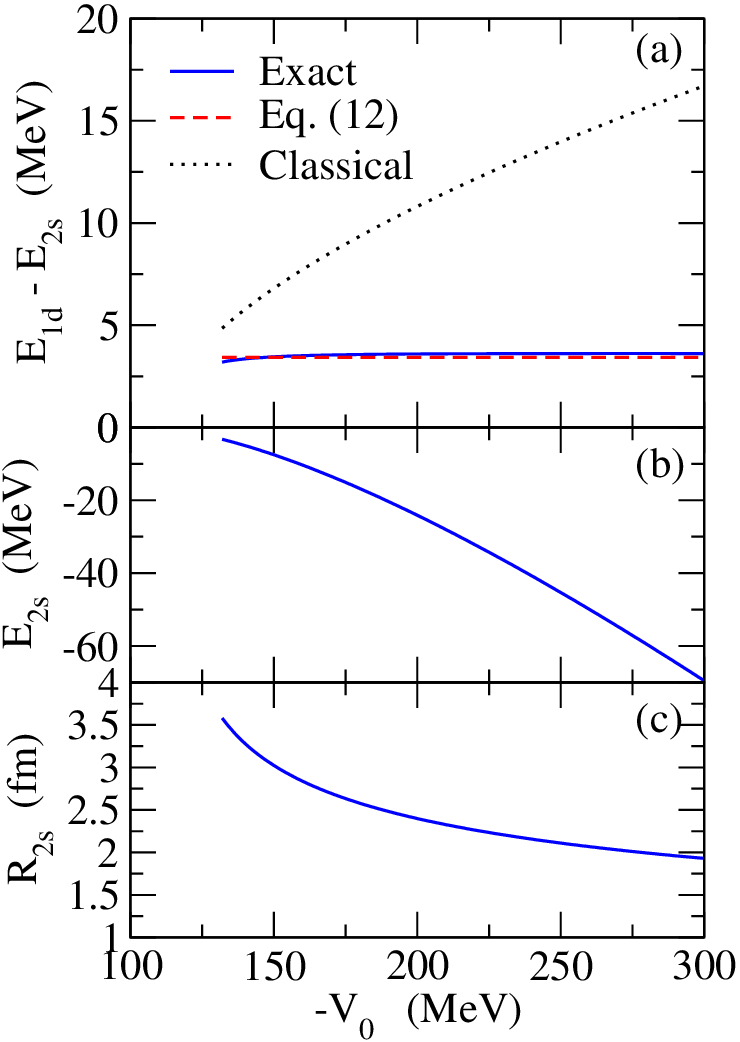}
\caption{(Color online)
(a) The energy difference between the $L=2$ (1d) and the $L=0$ (2s) states for a Gaussian 
potential $V(R)=V_0\,e^{-\beta R^2}$ for a $\alpha+\alpha$ system 
as a function of the depth parameter $-V_0$. 
The solid line is obtained by numerically solving the Schr\"odinger equation, while the 
dashed line is the result of the approximate formula given in Eq. (\ref{rowley}) 
based on the Bohr-Sommerfeld quantization rule. 
The dotted line denotes the energy difference for a classical rotor, evaluated with 
the root-mean-square (rms) intercluster distance for the 2s state shown in Fig. 8(c). 
(b) The energy of the $L=0$ (2s) state
as a function of the depth parameter $-V_0$ for the Gaussian potential. 
(c) The rms intercluster distance for the 2s state.  
}
\end{figure}

\medskip

\begin{acknowledgments}
We thank H. Tamura for useful discussions and for his continuing 
encouragement. 
This work was supported by the Japanese
Ministry of Education, Culture, Sports, Science and Technology
by Grant-in-Aid for Scientific Research under
the program number (C) 22540262.
\end{acknowledgments}

\appendix

\section{Rotational spectrum of a Gaussian potential}

In this Appendix, we numerically check the performance of 
the approximate formula given in Eq. (\ref{rowley}) for the rotational 
spectrum for a Gaussian potential, $V(R)=V_0\,e^{-\beta R^2}$. 
We consider here the $\alpha+\alpha$ system, and use the same geometry for the potential as that 
given in Eq. (\ref{gauss_aa}), {\it i.e.,} $\beta=0.22$ fm$^{-2}$. 
In this Appendix, we do not include the Coulomb interaction between the two $\alpha$ particles. 
Fig. 8(a) shows the energy difference between the $L=2$ state and the $L=0$ state for $2n+L=4$. 
Notice that without the Coulomb interaction the $L=2$ state is bound only for $V_0 < -131.5$ MeV. 
The solid line shows the numerical result of the Schr\"odinger equation, while the dashed line is 
obtained with Eq. (\ref{rowley}). One can see that Eq. (\ref{rowley}) works well, the deviation from 
the exact result being about $\pm$ 0.2 MeV for this system. 
The energy difference indeed depends on the depth parameter only weakly, while the energy and the root mean 
square (rms)
distance between the two $\alpha$ particles for the $L=0$ state 
varies from $-3.269$ MeV and 3.58 fm at $V_0=-132$ MeV to $-69.538$ MeV and 1.93 fm at $V_0=-300$ MeV, respectively 
(see Figs. 8(b) and 8(c)). 
As shown by the dotted line in Fig. 8(a), 
this is in marked contrast with the energy difference for a classical rotor, 
$\Delta E = 6\hbar^2/2\mu R^2$, where $\mu=M_\alpha/2$ is the reduced mass for the $\alpha+\alpha$ system and 
$R$ is the rms intercluster distance.

\end{document}